\documentclass{ieeetj}
\usepackage{cite}
\usepackage{amsmath,amssymb,amsfonts}
\usepackage{algorithmic}
\usepackage{graphicx,color}
\usepackage{textcomp}
\usepackage{xcolor}
\usepackage{hyperref}
\hypersetup{hidelinks=true}
\usepackage{algorithm,algorithmic}
\def\BibTeX{{\rm B\kern-.05em{\sc i\kern-.025em b}\kern-.08em
    T\kern-.1667em\lower.7ex\hbox{E}\kern-.125emX}}
\AtBeginDocument{\definecolor{tmlcncolor}{cmyk}{0.93,0.59,0.15,0.02}\definecolor{NavyBlue}{RGB}{0,86,125}}

\usepackage{graphicx}%
\usepackage{multirow}%
\usepackage{amsmath,amssymb,amsfonts}%
\usepackage{amsthm}%
\usepackage{mathrsfs}%
\usepackage{textcomp}%
\usepackage{manyfoot}%
\usepackage{booktabs}%
\usepackage{listings}%
\usepackage{lmodern}%
\usepackage{anyfontsize}%

\def\OJlogo{\vspace{-4pt}$<$Society logo(s) and publication title will appear here.$>$}
\def\seclogo{\vspace{10pt}$<$Society logo(s) and publication title will appear here.$>$}

\def\authorrefmark#1{\ensuremath{^{\textbf{#1}}}}

\begin{document}
\receiveddate{XX Month, XXXX}
\reviseddate{XX Month, XXXX}
\accepteddate{XX Month, XXXX}
\publisheddate{XX Month, XXXX}
\currentdate{XX Month, XXXX}
\doiinfo{XXXX.2022.1234567}

\markboth{}{Author {et al.}}

\title{LCDC: Bridging Science and Machine Learning for Light Curve Analysis}






\author{Daniel Kyselica\authorrefmark{1}, 
Tom\'{a}\v{s} Hrob\'{a}r\authorrefmark{2},
Ji\v{r}\'{i} \v{S}ilha\authorrefmark{2},
Roman \v{D}urikovi\v{c}\authorrefmark{1},
Marek \v{S}uppa\authorrefmark{1},
}
\affil{Department of Applied Informatics, Comenius University in Bratislava, 842 48 Bratislava, Slovakia}
\affil{Department of Astronomy, Physics of the Earth and Meteorology, Comenius University in Bratislava, 842 48 Bratislava, Slovakia}

\corresp{Corresponding author: Daniel Kyselica (email: daniel.kyselica@fmph.uniba.sk).}

\begin{abstract}
The characterization and analysis of light curves are vital for understanding the physical and rotational properties of artificial space objects such as satellites, rocket stages, and space debris. This paper introduces the Light Curve Dataset Creator (LCDC), a Python-based toolkit designed to facilitate the preprocessing, analysis, and machine learning applications of light curve data. LCDC enables seamless integration with publicly available datasets, such as the newly introduced Mini Mega Tortora (MMT) database. Moreover, it offers data filtering, transformation, as well as feature extraction tooling. 
To demonstrate the toolkit’s capabilities, we created the first standardized dataset for rocket body classification, RoBo6, which was used to train and evaluate several benchmark machine learning models, addressing the lack of reproducibility and comparability in recent studies. Furthermore, the toolkit enables advanced scientific analyses, such as surface characterization of the Atlas 2AS Centaur and the rotational dynamics of the Delta 4 rocket body, by streamlining data preprocessing, feature extraction, and visualization. These use cases highlight LCDC's potential to advance space debris characterization and promote sustainable space exploration.
Additionally, they highlight the toolkit's ability to enable AI-focused research within the space debris community.
\end{abstract}



\maketitle

\section{INTRODUCTION TO LIGHT CURVES}

Photometry is a technique used in astronomy to measure the light intensity or flux reflected or emitted by a celestial body. It is commonly used to obtain the dynamic (rotational) \cite{Vananti2023} and physical properties (size, albedo, shape, etc.) of artificial space objects such as satellites, rocket upper stages and space debris \cite{Friedman2021}. Brightness measurements are collected using astronomical telescopes. 

A light curve represents a series of brightness measurements of an object over time, with variations in brightness depending on the illumination conditions, viewing geometry, object’s shape, and attitude of the object. Additionally, significant brightness variations are often observed in space objects due to their rotation, leading to changes of several magnitudes, where the magnitude is a unit used in astronomy to describe the brightness of the object. It is a unit on a logarithmic scale, where five magnitudes represent a brightness change of 100 times. The magnitude is measured in the reverse direction, where the smaller number represents the higher brightness. Photometric light curves can contain magnitudes in arbitrary units (e.g. instrumental magnitudes), or they can contain calibrated brightness values when reported, as standard magnitudes calibrated to a specific photometric system using photometric catalogues \cite{Tonry2018}.

\subsection{LIGHT CURVE ANALYSIS}
For rotating space objects, the apparent (synodic) rotation period is routinely estimated using various techniques, such as Fourier analysis, Phase Dispersion Minimization \cite{Schwarzenberg2009}, etc. Once the synodic period is determined, the light curve can be folded using the apparent rotation period, allowing for a much better analysis of the signal's complexity and enabling the extraction of the mean brightness value. Usually, a folded light curve is fitted with the Fourier series \cite{SILHA20202018}.

\subsection{PHOTOMETRIC CATALOGUES}
Currently, positions of more than 27,000 artificial space objects are regularly collected, and the orbits of the objects are catalogued \cite{stace2025track}. However, these observations do not allow for the tracking and archiving of brightness measurements. As a result, several research institutions are working to create a brightness catalogue of artificial space objects. Examples include the Ukrainian Light Curve Database and Atlas of Odessa University \cite{Koshkin2017}, the Light Curve Catalogue of the Astronomical Institute of the University of Bern \cite{SILHA2018844}, the Mini-MegaTORTORA Database (MMT) of Kazan Federal University \cite{karpov2016massive}, and the Space Debris Light Curve Database (SDLCD) of Comenius University \cite{SILHA20202018}. Of these, the latter two catalogues are the only ones publicly available to the community.

\subsection{MMT DATABASE}
\begin{figure}[ht!]
    \centering
    \includegraphics[width=\linewidth]{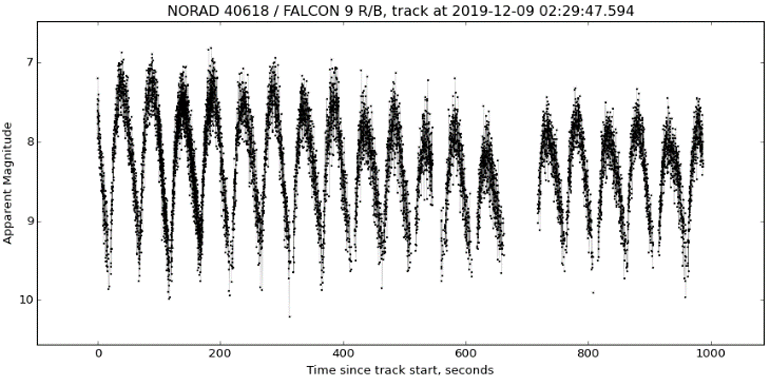}
    \caption{Example of Falcon 9 light curve from MMT database \cite{mmtWeb}.}
    \label{fig:mmt_example}
\end{figure}

The largest source of the publicly available light curves dataset is the Russian Mini Mega Tortora (MMT) database \cite{karpov2016massive,mmtWeb} a wide-field monitoring system operated by the Special Astrophysical Observatory of the Russian Academy of Sciences. 
As of the 7$^\text{th}$ of January 2025,
the database contained 15, 825 objects and 534, 604 tracks (or light curves),  with new records being added daily. The light curves of objects, identified by their Norad ID, are stored in a sequence of measurements containing the time, standard magnitude, and phase angle. 
Figure~\ref{fig:mmt_example} presents a sample light curve of the Falcon 9 rocket body. The database is widely used for training and validation 
for object characterization and attitude determination, as described in Section~\ref{sec:ml_intro}. 

\subsection{SDLCD DATABASE}
\begin{figure}[ht!]
    \centering
    \includegraphics[width=\linewidth]{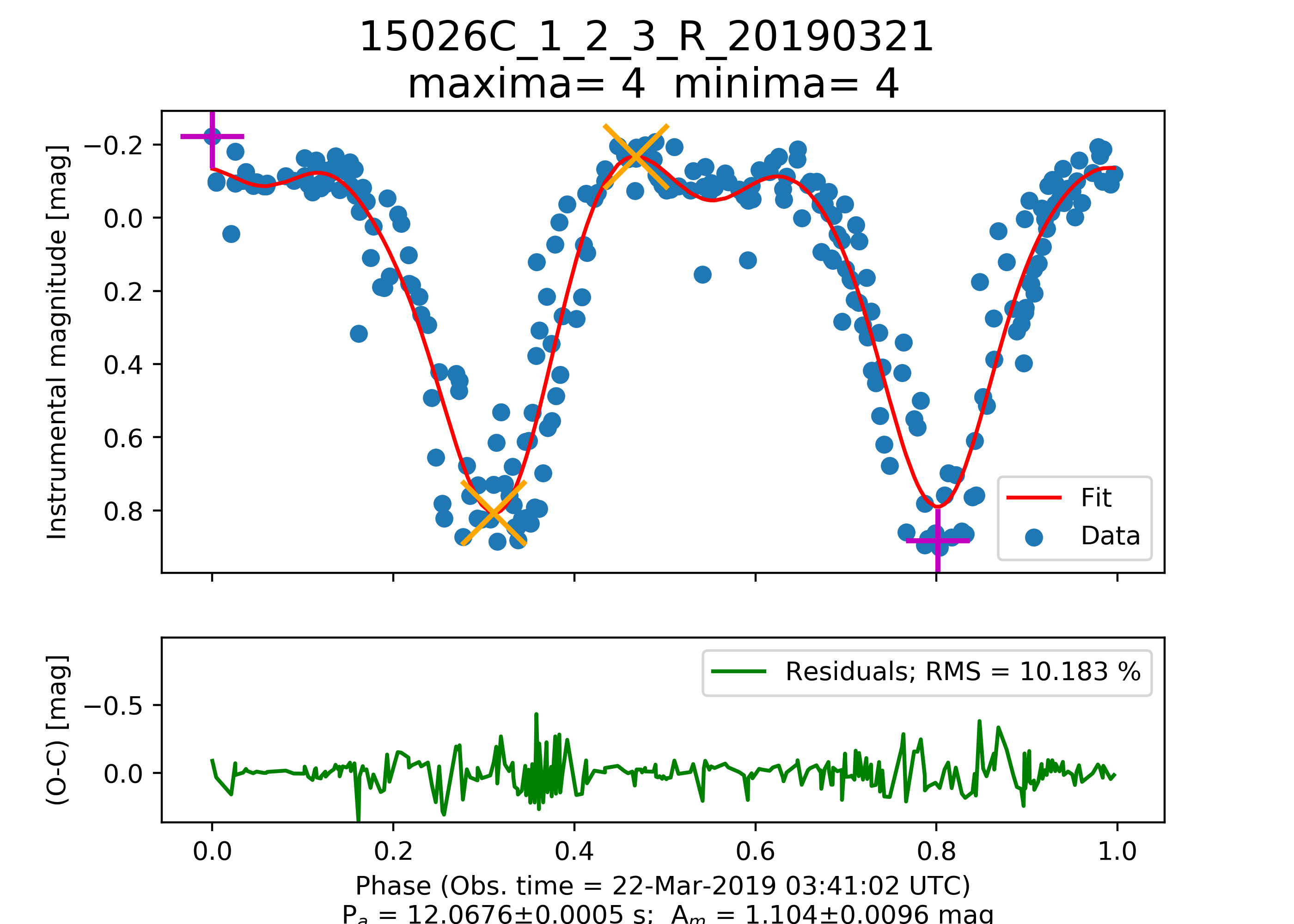}
    \caption{Reconstructed rotation phase of spared Ariane 5 upper stage published in the SDLCD. For this specific light curve, the apparent rotation period of the object was found to be 12.1 seconds. Source: \url{https://www.sdlcd.space-debris.sk}}
    \label{fig:sdlcd_example}
\end{figure}

Complementary to the MMT catalogue is the SDLCD, which also contains publicly available photometric data, including original preprocessed light curves, extracted apparent periods, and constructed rotation phases (\cite{SILHA20202018}). As of the 22$^\text{nd}$ of January 2025, the SDLCD contains a total of 2,224 light curves of 791 individual objects, including 212 spacecraft, 521 rocket bodies, and 58 other space debris objects that covered the observation period from January 2016 up to June 2024. Compared to MMT, brightness values in SDLCD are reported in so-called instrumental magnitudes, which are arbitrary units not linked to an object's physical brightness, although the brightness variations are preserved. The database can be accessed online at \url{https://www.sdlcd.space-debris.sk}. 
Alternatively, the entire dataset is available upon request from the authors \cite{SILHA20202018}. 
Figure~\ref{fig:sdlcd_example} shows an example of the reconstructed rotation phase and its Fourier fit for the upper stage from European rocket launcher Ariane 5. 

\section{MACHINE LEARNING ON LIGHT CURVES}\label{sec:ml_intro}

From a space object light curve various information can be extracted. A typical task involves the determination of attributes such as shape, size, and rotation. Machine learning models are dependent on their training data, and the MMT database has become widely used because of its large size and accessibility. Machine learning on light curves can help resolve many issues in this area and was adopted in several domains described in this section.
Within this research domain, preprocessing, evaluation, and normalization strategies differ greatly between studies, making comparison of different methods very difficult and oftentimes impossible. 
Although this section highlights only a subset of works in the subfield, it is evident that there are substantial differences in the preprocessing, evaluation, and normalization strategies across studies. These inconsistencies make direct comparisons between methods highly challenging, if not impossible, emphasizing the need for a standardized approach to enable meaningful evaluation. 

\subsection{OBJECT CHARACTERIZATION}
\label{sec:ml_intro-object_characterization}

One of the first uses of the database for machine learning was by the authors in \cite{furfaro2018space}. They have set up a classification task using a custom 1-D convolutional network with 3 different target classes, namely: rocket bodies, debris, and satellites. A sequence of the first 500 measurements from the light curve was used as an input, with the reported test accuracy of around 75\%. 

A similar approach was used in \cite{linares2020space} with the same input sequence strategy and 1-D convolutional neural network (CNN)  achieved a comparable accuracy of 75\%, but on a different set of target classes.
Rather than relying solely on the magnitude information, the authors added the second channel with the phase angle data to the input thus using more of the available information. 
The uniform size of the input sequences was achieved by truncation and zero padding of the original light curve to 1200 points, corresponding to approximately 20 minutes of observation time. Together with the pre-training on the synthetic data, the 1-D CNN network was able to correctly classify objects from 8 classes (7 rocket body classes + bow-wing class) with 90\% accuracy. Likewise in the work \cite{kerr2021light}, 1-D CNN and LSTM-based models were trained to classify objects within GEO orbit into categories based on their shapes and size. To ensure uniform temporal sampling, the inputs were resampled to a consistent frequency and,
as in the previous studies, padded and truncated to a uniform size. 



Instead of padding and truncation, the authors \cite{yao2021basic} used a folding technique to standardize the input while minimizing information loss. Light curves were folded by their rotational period, producing a new sample with 200 measurements each. 
This preprocessing technique, combined with data augmentations such as phase translation and noise addition, enabled a 1-D CNN to achieve 85\% accuracy in a classification scenario involving 5 rocket bodies. Building on prior work, the authors in \cite{kyselica2022astronomical} divided light curves into individual rotational periods and rescaled them into uniform size, resulting in a significantly expanded training dataset, which was used to train the 1-D version of ResNet20~\cite{he2016deep} network for classification of three rocket bodies, achieving 84\% accuracy. 

A contemporary approach to handling large sequences is using the Transformer \cite{vaswani2017attention} module. Astroconformer, a Transformer-based model introduced \cite{shrive2024classifying}, was originally designed for 
surface gravity estimation from stellar light curves processing, yet it can be easily adapted for classification tasks.

\subsection{ATTITUDE DETERMINATION}

Machine learning approaches are also being used to extract rotational information from the light curve.  Recurrent neural networks (RNNs) can be used to estimate an object's rotational state \cite{alhajeri2021machine}, but often this problem is reinterpreted to classification. In \cite{Agathanggelou2022}, the authors analyzed various techniques such as Random Forests and RNNs for the classification of objects into tree categories: active, inactive and decayed. The CNNs are commonly used in the former way in works \cite{badura2022convolutional}, \cite{kerr2021light} where authors use different classifying criteria. However, CNNs do not perform well in the case of state estimation. An interesting approach was presented in \cite{alhajeri2021machine} where the best-performing model for state estimation included a Feedforward neural network together (representing a non-linear process) with a physical-based computation module. 
The researchers \cite{ISOLETTA2024} created an architecture that exploits the Lomb-Scargle Periodogram \cite{vanderplas2018understanding} and the Phase Dispersion Minimization algorithm \cite{stellingwerf1978period} for object classification into three categories: Spin-stabilized, tumbling and uncertain with 90\% accuracy.

\subsection{OTHER DOMAINS}

Information extraction from noisy measurements can be enhanced using the right preprocessing techniques like Fourier series expansion, as examined by \cite{gallucci2022light}, and wavelet transformation, applied by \cite{qashoa2023classification} etc. However, 
distinguishing objects with similar physical attributes, especially rocket bodies, continues to pose a major challenge.

The authors in \cite{Li_2024} addressed the challenge of detecting potential contaminants in light curves. Stellar contamination arises when an object passes in front of a star, causing its light to be misattributed. In contrast, cloud contamination occurs when an object is partly obscured by clouds. The researchers achieved the best test accuracy using a custom CNN for stellar contamination and an SVM for cloud contamination of 92\% and 97\%, respectively.

Another study \cite{badura2024identifying} focuses on using machine learning to narrow down initial orbit determination search spaces to six specific cislunar orbital families. Synthetic data was generated for training using the Digital Imaging and Remote Sensing Image Generation (DIRSI$^{\text{TM}}$) engine \cite{goodenough2012dirsig}, covering six orbital families. The trained model achieved an 87.5\% test accuracy when combined light curves captured from space-based L1 point and the face of the moon were used.

A Modern approach in dealing with large sequences today is using the Transformer \cite{vaswani2017attention} module. In the stellar light curve community transformer models such as Astroconformer \cite{shrive2024classifying}, can achieve outstanding results to the more traditional network if presented with a sufficient amount of training data. 
Despite being designed for processing stellar light curves for surface gravity estimation, this pretrained model can be easily adapted for the classification of space debris light curves.

\section{LIGHT CURVE PROCESSING TOOLS}
\label{sec:tools}

Light curves are extensively utilized across various fields of astrophysics, with stellar light curves being the most prominent. This necessitates the development of tools for their analysis and interpretation. In this section, we examine three Python libraries and tools designed to streamline working with light curves.

\subsection{PROCESSING TOOLBOX FOR PYTHON - LIGHT-CURVE}

The feature extraction library called \texttt{light-curve} \cite{malanchev2021anomaly} is implemented in both Python and Rust programming languages to ensure high computational efficiency and performance. Within this library, the majority of the classes are designed to implement a range of feature extractors, which are essential tools for the classification and characterization of astrophysical sources based on their light curves.

The library contains multiple possibilities for function fitting like the Villar function, Bazin function, Linexp function etc. Moreover, the library gives access to computing various statistics including Unbiased Anderson–Darling normality test statistic, excess kurtosis of magnitude, maximum slope between two sub-sequential observations and more. The complete list of functions is presented in the official documentation~\cite{lightcurve2024pypi}.

\subsection{ASTROBASE}

Python package Astrobasev~\cite{wbhatti_astrobase} has been developed primarily for analysis purposes of stellar light curves. It is capable of reading light curves in various formats including Kepler mission format \cite{nasa2024kepler} and TESS (Transiting Exoplanet Survey Satellite) \cite{nasa2024tess} format.
It features various period-finding algorithms, batch-processing tools for extensive light curve datasets, and a web application for analyzing and categorizing light curves based on stellar variability. This package can be used to work with any light curves in the correct format of three sequences: measured magnitudes, timestamps and measurement error.
The functionality is divided into the following modules:
\begin{itemize}
    \item \textit{lcfit} - Module for fitting the input data with selected functions like Fourier series, cubic spline, Savitzky-Golay smoothing filter~\cite{savitzky1964smoothing}, Legendre function~\cite{dunster2009legendre}, and more;
    \item \textit{lcmath} - Offers several essential tools for light curve analysis, including phasing, sigma-clipping, and gap identification and interpolation;
    \item \textit{lcmodels} -  Includes several light curve models for variable stars, facilitating initial fits to differentiate variable types and construct light curves for recovery simulations;
    \item \textit{varbase} - Contains functions for light curve variability determination, fitting functions, masking signals, autocorrelation, etc.;
    \item \textit{plotbase} - Plotting functionality for light curves;
    \item \textit{lcproc} - Enables batch processing of multiple samples in parallel.
\end{itemize}

Other functionality includes visualization and utilities for variable star classification.

\subsection{LIGHTCURVES BY SWAGNER-ASTRO}

A common approach by researchers is to create smaller libraries to tackle a specific problem. For instance, \textit{Lightcurves} package \cite{wagner2021statistical} was created to characterize flares, using Bayesian block representation  \cite{scargle2013studies} and the HOP algorithm from \cite{meyer2019characterizing}.

\section{CONTRIBUTION}

As discussed in Section~\ref{sec:ml_intro}, light curves are ideal for machine learning as demonstrated by multiple studies. However, one of the requirements of solid scientific research is that a work needs to be reproducible, which is not the case in many instances,
as discussed in Section \ref{sec:ml_intro-object_characterization} such as \cite{yao2021basic}, \cite{furfaro2018space} etc. This can be attributed to variations in filtration and preprocessing, even when the source data is identical, making result reproduction nearly impossible. Although some studies provide details about preprocessing, the code or algorithms used are often omitted. The available Python libraries for light curve processing presented in Section~\ref{sec:tools} are focused primarily on processing a single stellar light curve. Moreover, they lack the capability of dealing with large datasets for machine learning models. In this work, we tackle these problems by:

\begin{enumerate}
    \item Introducing a standardized and reproducibly pre-processed Snapshot of the MMT database available on the HugginFace website with regular updates, making data easy to access and lowering the internet traffic on MMT website \url{https://huggingface.co/datasets/kyselica/MMT_snapshot}.
    \item Introducing LCDC: Light Curve Dataset Creator toolkit written in Python, whose application is suitable for both machine learning and the astronomical community.
    \item Introducing RoBo6: Standardized MMT Light Curve Dataset For Rocket Body Classification. The dataset enables model comparison, facilitates consistent evaluations and accelerates advancements in this research area.
\end{enumerate}

The capability of light curve analysis over large databases 
can be crucial for understanding an object's physical (surface characteristics) \cite{Hejduk2011} and dynamical properties (rotation). Knowing these attributes helps reduce undesirable effects on the environment and the overall removal of objects in the future. LCDC tool together with \texttt{MMT\_snapshot} is ideal for mentioned scientific analysis.

The subsequent sections of this article are organized as follows: Section~\ref{sec:lcdc} introduces the LCDC toolkit, describing its architecture, functionality, and data preprocessing capabilities for large light curve datasets. Moreover, the \texttt{MMT\_snapshot} dataset is introduced in the same Section~\ref{sec:LCDC-input}. 
The following sections showcase experiments and use cases of the LCDC toolkit for both machine learning and scientific purposes. Section~\ref{sec:robo6} is devoted to the creation of the first standardized rocket body dataset RoBo6, and its evaluation with different machine learning models. Section~\ref{sec:sur_char} and~\ref{sec:rot_prop} contain scientific analyses of specific space objects, namely surface characteristics of ATLAS 2AS rocket body and rotational dynamics evolution of Delta 4 rocket body. 

\section{LCDC TOOLKIT}
\label{sec:lcdc}

We introduce LCDC: an open-source dataset creator toolkit for machine learning and statistical analysis of light curves. It allows 
for simple and efficient data handling and preprocessing of large light curve datasets suited for both scientific analysis and training of machine learning models.

The toolkit, implemented in Python, is structured around a \texttt{DatasetBuilder} module, \texttt{datasets.Dataset} module from the Python \texttt{datasets} package maintained by HuggingFace~\cite{datasets2024} and the Pyarrow Python library for handling Apache Arrow format \cite{apache2025arrow}. The source code is released publicly under the terms of the MIT license on GitHub (\url{https://github.com/lcdc-develop/lcdc}).

\subsection{INITIALIZATION}

To create a dataset, the main \texttt{DatasetBuilder} class is initialized with arguments described in Tab.\ref{tab:DatasetBuilder-arguments}. The optional arguments control the initial selection of objects, either by NORAD indices defined in the \texttt{norad\_ids} or by regular expression of particular classes defined in  \texttt{regexes} and \texttt{classes} arguments.  

\begin{table}[h]
    \centering
    \caption{Arguments of the DatasetBuilder class.}
    \begin{tabular}{p{1.7cm} | p{5cm} }
        \toprule
        \textbf{Argument name} & \textbf{Description} \\ 
        \midrule
        \textit{directory} & Path to data directory \\ 
        \textit{classes}  & Class names (Optional). \\ 
        \textit{regexes} & Regular expressions corresponding to classes (Optional). \\
        \textit{norad\_ids} & List containing NORAD indices of objects of interest (Optional). \\
        \bottomrule
    \end{tabular}
    \label{tab:DatasetBuilder-arguments}
\end{table}

\subsection{INPUT}
\label{sec:LCDC-input}
LCDC expects the data to be stored as one or more \texttt{.parquet} files in a single directory. All records in a \texttt{.parquet} file must contain the following columns: 

\begin{table}[h]
    \centering
    \caption{Description of light curve .csv columns.}
    \begin{tabular}{p{1.7cm} | p{5cm}}
        \toprule
        \textbf{Column} & \textbf{Description} \\ 
        \midrule
        \textit{id} & Identification number of the record. \\
        \textit{norad\_id} & NORAD identification number of the object.\\
        \textit{period} & Apparent rotational period in seconds.\\
        \textit{timestamp} & Start of the observation timestamp in the format '\%Y-\%m-\%d \%H:\%M:\%S' \\
        \textit{name} & Name of the object.\\
        \textit{variability} & Variability: 0 (nonvariable), 1 (aperiodic), 2 (periodic).\\
        \textit{time} & A list of observation times. \\
        \textit{mag} & A list of measured standardized magnitudes to 1000 km distance. \\
        \textit{phase} & A list of measured phase angles in degrees.\\
        \textit{distance} & A list of measured, altitudes in km. \\
        \textit{filter} & A list of filters used: 0 (unknown), 1 (clear), 2 (pol), 4 (V), 8(R), 16(B).\\
        \bottomrule
    \end{tabular}
    \label{tab:csv-columns}
\end{table}

\subsubsection{\texttt{MMT\_SNAPSHOT}}

Data from the MMT database are publicly available and can be downloaded in bulk, however, its format is not easy to use as it consists of a large number of text files with different internal formats. To address this problem and to limit the server traffic, a snapshot of the database called \texttt{MMT\_snapshot} was created and is regularly updated in the correct format and published on Hugging Face website \url{https://huggingface.co/datasets/kyselica/MMT_snapshot}. Data is stored in the format described in the previous section suitable for LCDC, thus seamlessly enabling it to work with the whole MMT database.

\subsection{OUTPUT}

Intermediate states of \texttt{DatasetBuilder} can be stored on a disk in \texttt{.parquet} file using \texttt{to\_file(path)} method. Another possibility is invoking \texttt{build\_dataset(split\_ratio=None)} method producing one or two \texttt{datasets.Dataset} 

\subsection{PREPROCESSING}

Preprocessing is an essential phase in ensuring data quality when handling large datasets. To ensure the resulting dataset is of acceptable quality, a range of filtering criteria, transformations and splitting operations can be applied to the initial data.  All methods are available in the \texttt{preprocessing} and \texttt{stats} modules. Desired operations are specified in an ordered list and applied to data using the \texttt{DatasetBuilder.preprocess(<funcs>)} method.

\subsection{FILTERING}

Large datasets contain low-quality samples, which decreases their overall quality. Therefore, the LCDC package offers several ways of filtering out undesirable examples:

\begin{itemize}
    \item \texttt{FilterByStartDate(Y,M,D,h,m,s)} 
    
    The first observation from the sample must not be taken before the set time.
    \item \texttt{FilterByEndDate(Y,M,D,h,m,s)}
    
    The first observation from the sample must not be taken after the set time.
    \item \texttt{FilterByNorad(norad\_list)}
    
    The NORAD ID of the sample must be present in the \texttt{norad\_list}.
    \item \texttt{FilterByPeriodicity(types)}
    
    The sample must be marked by one of the variability \texttt{types}: aperiodic, periodic or non-variable. 
    \item \texttt{FilterByMinLength(length,step: Optional)}
    
    The sample must contain at least \texttt{length} measurements. If the \texttt{step} parameter is set, the sample is first resampled by its value.
    \item \texttt{FilterFolded(k,threshold)}
    
    The sample is folded by its apparent rotational period, producing a new sample spanning over one period with \texttt{k} points. At least \texttt{threshold} of the new sample needs to be covered by measurements.

\end{itemize}

\subsection{TRANSFORMATIONS}

Often in the raw data, the information can be invisible due to the noise present. To extract information from the data or transform the data into the correct form, transformation is used. Currently, there are two possibilities implemented in the \texttt{preprocessing} module: 

\begin{itemize}
    \item \texttt{Fold}
    
    The sample is folded by its apparent rotational period, producing a new sample that spans over one period.
    \item \texttt{ToGrid(sampling\_frequency, size)}
    
    The sample is resampled with the \texttt{sampling\_frequency} and trimmed or padded to the desired \texttt{size}.
\end{itemize}

\subsection{SPLIT METHODS}

Some machine learning algorithms rely on the grid-like structure of the input. To ensure the uniform size of the input, padding, truncation and splitting operations can be applied. The splitting operation in contrast to truncation does not result in information loss. Moreover, for analysis of rotating objects it is natural to think in terms of individual rotational periods. Therefore, multiple splitting strategies can be used to achieve desired results:

\begin{itemize}
    \item \texttt{SplitByGaps(max\_length=None)}
    
    Splits the sample into two parts, where the difference between two consecutive measurements is greater than its apparent rotational period. When \texttt{max\_length} parameter is set, the sample with a smaller length than this will not be split. 

    \item \texttt{SplitByRotationalPeriod(multiple=1)}

    A sample is split at \texttt{multiple} of its apparent rotational period.

    \item \texttt{SplitBySize(max\_length,uniform=False)}

    Splits sample by the \texttt{max\_length}. If the \texttt{uniform} flag is set for the uniforms splitting.
    
\end{itemize}

\subsection{STATISTICS}

Certain properties of an input sample that cannot be directly observed become apparent when the appropriate statistical methods are applied. The \texttt{stats} module provides implementations of common statistical techniques and feature extractors for light curves, including:
\begin{itemize}
    \item \texttt{Amplitude}

    Computes the amplitude of a sample as the difference between maximal and minimal observed magnitude.
    
    \item \texttt{MediumTime} 

    Computes the time average of the observations of the sample.

    \item \texttt{MediumPhase}

    Computes the phase average of the observations of the sample.

    \item \texttt{FourierSeries(order,fs=True,
    amplitude=True)}

    Computes the Fourier series of \texttt{order} \cite{weissteinfs} of the input signal, described in Equation~\ref{eq:fs}. The flags \texttt{fs} and \texttt{amplitude} can be set to store the computed coefficients and amplitude computed from the reconstructed signal as shown in Equation~\ref{eq:fs_amp}, respectively.

    \begin{equation}\label{eq:fs}
    f(t) \approx \frac{1}{2}a_0 + \sum_{n=1}^{N}( a_n \cos(2\pi \frac{n}{P} t) + b_n \sin(2\pi \frac{n}{P} t) ).
    \end{equation}
    \begin{equation}\label{eq:fs_amp}
    A = \text{max}_t f - \text{min}_t f
    \end{equation}

    \item \texttt{ContinousWaveletTransform(wavelet,step,
    length,scales)}

    Discrete version of Continuous wavelet transform \cite{wavelet_book} is computed using Equation~\ref{eq:cwt} where $x(t)$ is the original signal, $1 \leq a \leq \texttt{scales}$ and $0 \leq b < length; b = i \texttt{ step}$, are the scale and translation parameters that
    are used to stretch and shift the wavelet function $\psi(t)$. The \texttt{wavelet} parameter is a base function compatible with the 
    PyWavelets package \cite{lee2019pywavelets}.  

    \begin{equation}\label{eq:cwt}
    X_{\psi}(a, b) = \frac{1}{|a|^{1/2}} \sum_{t=0}^{T} x(t) \psi(\frac{m-b}{a}).
    \end{equation}\label{eq:dcwt}
    
\end{itemize}

\subsection{VISUALISATION}

\begin{figure*}
    \centering
    \includegraphics[width=.8\linewidth]{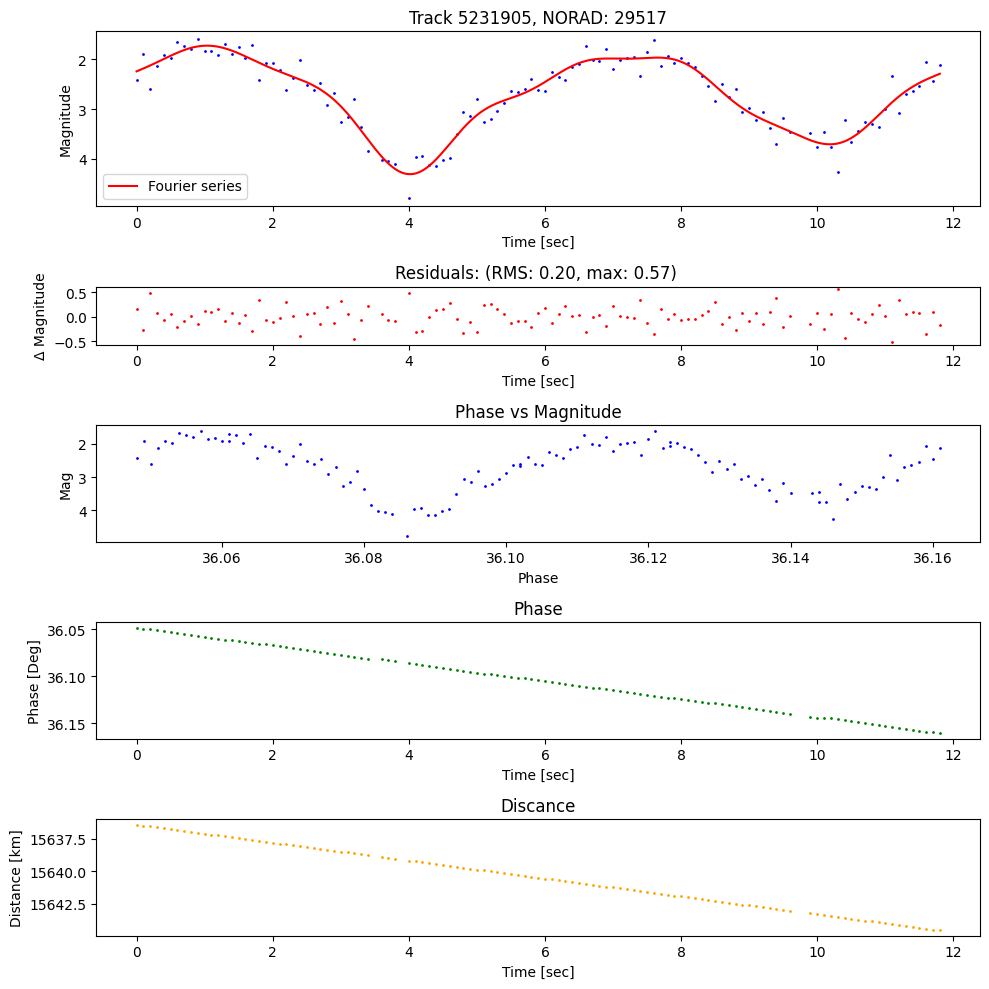}
    \caption{Visualisation of a part of the light curve with ID 5231905 of a CZ-3B rocket body with NORAD ID 29517.  }
    \label{fig:visualisation}
\end{figure*}

Data visualization often aids in identifying outliers and gaining a deeper understanding of system behaviour. The \texttt{utils} module provides visualization capabilities, including plotting a sample's magnitude over time, phase over time, distance over time, and Fourier series reconstruction with residuals, provided all required information is present in the data sample, as illustrated in Figure~\ref{fig:visualisation}.

\section{USE CASE I: ROBO6 DATASET}
\label{sec:robo6}

As described in Section~\ref{sec:ml_intro-object_characterization}, object characterization is one of the areas where machine learning is actively used to solve this problem.

Prior work has predominantly focused on classifying objects such as rocket bodies, debris, or satellites based on their light curves. However, a more practically relevant challenge lies in distinguishing between objects with a similar shape, such as different types of rocket bodies. To the best of our knowledge, no standard benchmark currently exists for addressing this specific classification problem. 

To address this gap, using the LCDC package and MMT database we curated a standardized dataset comprising six common rocket body populations: CZ-3B, Atlas 5 Centaur, Falcon 9, H-2A, Ariane 5 and Delta 4. 
The dataset is divided into a train set of 5,676 samples and a test set of 1,404 samples, with further details provided in Table \ref{tab:dataset1}.

\begin{table}[ht]
    \centering
    \caption{Number of samples per split and class.}
    \begin{tabular}{lrr}
    \toprule
    \textbf{Class} &  \textbf{Train} &  \textbf{Test} \\
    \midrule
    \textbf{Ariane 5}        &    660 &   173 \\
    \textbf{Delta 4}         &    233 &    70 \\
    \textbf{CZ-3B}           &   2266 &   548 \\
    \textbf{Atlas 5 Centaur} &   1029 &   247 \\
    \textbf{H-2A}            &    623 &   150 \\
    \textbf{Falcon 9}        &    865 &   216 \\
    \bottomrule
    \end{tabular}
    \label{tab:dataset1}
\end{table}

Each sample is characterized by the following fields: \textit{label}, \textit{ID}, \textit{part} number, \textit{period} (in nanoseconds), \textit{mag}, \textit{phase} and \textit{time}. The last three fields refer to file paths that contain additional metadata, such as standard magnitude, phase angle, and time measurements. 
During preprocessing, samples can be divided into subsamples, with their sequence order recorded in the \textit{part} field.

Since many machine learning models, such as CNNs, require grid-structured data, each sample needs to be resampled onto a uniformly spaced grid with a manageable length. To convert the data into this format and to retrain as much information as possible, a series of filtering, splitting and resampling operations was performed over each sample. Standard normalization using mean and standard deviation was also applied.

By enabling easy comparison across models and providing robust training and testing splits, RoBo6 facilitates consistent evaluations and accelerates advancements in this research area. Models trained on the dataset demonstrated its suitability for both traditional and transformer-based architectures, with Astroconformer achieving the highest performance as described in Sec.~\ref{sec:robo6-results}. We hope this dataset fosters advancements in space object classification and promotes sustainable practices in space exploration.

The dataset is publicly available on the HuggingFace platform (\url{https://huggingface.co/datasets/kyselica/RoBo6}) and was created from the \texttt{MMT\_snapshot} described in Sec.~\ref{sec:LCDC-input}.

\subsection{DATASET CREATION}

To create a dataset of high quality a preprocessing operation needs to be applied to the raw MMT data. This can be achieved using the new LCDC toolkit. 

\textit{Splitting:}
The original data contains light curves that often span extended periods. As CNN models require a grid input structure these areas would be translated into long sequences filled with zeros. The analysis has shown, that by using \texttt{SplitByGaps}, 95\% of the new samples were found to be shorter than 1,000 seconds. Therefore, this splitting operation was used following by \texttt{SplitBySize} with \texttt{max\_length} parameter set to $1000$ seconds and with the $uniform$ flag set, to split the remaining longer samples achieving a dataset with samples of size $\leq 1000 \text{ sec}$.

\textit{Filtering:}
Low-quality samples are filtered based on two empirically determined criteria consistent with
values commonly used in prior studies, such as \cite{furfaro2018space}. The first criterion sets the minimal number of measurements in a sample to $100$. The second criterion accepts only samples with at least $75\%$ coverage of its folded version. Using \texttt{FilterByMinLength(length=100)} followed by  \texttt{FilterFolded(k=100,threshold=0.75)} performs desired filtration.

\textit{Transformations:}
To standardize the dataset, all samples were rescaled and zero-padded to a fixed length of 10, 000 points using \texttt{ToGrid} functionality with the \texttt{size} of 10, 000 and the \texttt{sampling\_frequency} 10Hz based on the analysis of the time intervals between consecutive measurements. 

\subsection{EVALUATION STRATEGY AND METRIC SELECTION}

Selecting an effective evaluation method is essential for the precise assessment of model performance. A viable strategy involves partitioning data by objects into train and test sets to gauge the model's competence in classifying new objects. Nevertheless, this method is challenging with multiple classes due to the limited object count and a significantly imbalanced light curve distribution, potentially resulting in an inadequately representative test set and insufficient samples for reliable evaluation. As shown in Tab.~\ref{tab:dataset2}, the disparity in data distribution is particularly evident in the case of the CZ-3B rocket, where just 14 objects account for more than $60\%$ of all light curve tracks.

\begin{table}[ht]
    \centering
    \caption{Number of tracks per CZ-3B rocket divided based on the number of tracks.}
    \begin{tabular}{c|c|c|c}
    \toprule
    \textbf{Track count} & \textbf{\# objects} & \textbf{\# tracks} & \textbf{Coverage} \\
    \midrule
    326 - 391& 1 & 391 & 7.87 \% \\
    261 - 326& 2 & 546 & 10.99 \% \\
    196 - 261& 6 & 1303 & 26.22 \% \\
    131 - 196& 5 & 782 & 15.74 \% \\
    66 - 131 & 10 & 907 & 18.25 \% \\
    1 - 66   & 63 & 1040 & 20.93 \% \\
    \bottomrule
    \end{tabular}
    \label{tab:dataset2}
\end{table}

In real-world scenarios, the same object can be observed multiple times and must be consistently identified. To simulate this, the dataset can be split by track ID, ensuring all samples derived from the same light curve during preprocessing are assigned to the same set. Stratified splitting is employed to maintain consistent class distributions across the training and testing sets.

Given the imbalanced nature of the dataset, accuracy alone can be a misleading metric, since strong performance on classes with many samples may overshadow poor performance on underrepresented classes.
To address this issue, the F1 macro score was chosen as the primary evaluation metric, as it takes into account the cardinality of all classes and thus provides a more balanced assessment of model performance across the dataset.

\subsection{MODEL TRAINING AND PERFORMANCE EVALUATION}\label{sec:robo6-results}

To demonstrate the utility of the dataset, five selected models \cite{allworth2021transfer},~\cite{he2016deep},~\cite{furfaro2018space},~\cite{yao2021basic} and~\cite{pan2024astroconformer} were trained on it. Each model was trained for 50 epochs using the Adam optimizer with a learning rate of $0.001$. Training parameters specified in the respective publications were followed; in cases where parameters were not provided, default settings were used. The complete list of parameters and further preprocessing details can be found in  Tab.~\ref{tab:params}.
The results, summarized in Tab.~\ref{tab:results_acc_f1} and \ref{tab:results_prec_recall}, align closely with those reported in the respective publications. Astroconformer, despite being developed for stellar light curves, proved to be the best-performing model in this benchmark comparison.
\begin{table}[htbp]
    \centering
    \caption{Models performance: Accuracy and F1.}
    \label{tab:results_acc_f1}
    \begin{tabular}{c|c|c}
\toprule
\textbf{Model} & \textbf{Accuracy} & \textbf{F1} \\
\midrule
\textbf{ALLWORTH}~\cite{allworth2021transfer}   & 0.559 ± 0.044 & 0.478 ± 0.038 \\
\textbf{ResNet}~\cite{he2016deep}              & 0.694 ± 0.023 & 0.600 ± 0.034 \\
\textbf{FURFARO}~\cite{furfaro2018space}       & 0.628 ± \textbf{0.009} & 0.552 ± \textbf{0.013} \\
\textbf{YAO}~\cite{yao2021basic}              & 0.672 ± 0.017 & 0.604 ± 0.023 \\
\textbf{Astroconf.}~\cite{pan2024astroconformer} & \textbf{0.725} ± 0.011 & \textbf{0.684} ± 0.015 \\
\bottomrule
    \end{tabular}
\end{table}
\begin{table}[ht]
    \centering
    \caption{Models performance: Precision and Recall.}
    \label{tab:results_prec_recall}
    \begin{tabular}{c|c|c}
\toprule
\textbf{Model} & \textbf{Precision} & \textbf{Recall} \\
\midrule
\textbf{ALLWORTH}~\cite{allworth2021transfer}   & 0.491 ± 0.033 & 0.531 ± 0.024 \\
\textbf{ResNet}~\cite{he2016deep}              & \textbf{0.738} ± 0.026 & 0.584 ± 0.033 \\
\textbf{FURFARO}~\cite{furfaro2018space}       & 0.570 ± 0.017 & 0.552 ± \textbf{0.013} \\
\textbf{YAO}~\cite{yao2021basic}              & 0.622 ± 0.029 & 0.601 ± 0.020 \\
\textbf{Astroconf.}~\cite{pan2024astroconformer} & 0.702 ± \textbf{0.010} & \textbf{0.677} ± 0.019 \\
\bottomrule
    \end{tabular}
\end{table}

%
The hyperparameters used for the training of the respective models are detailed in  Table~\ref{tab:params}.  In an attempt to reproduce the published models as closely to their published state as possible, a specific preprocessing step was incorporated specifically for these models:

\begin{itemize}
    \item \textbf{ALLWORTH}~\cite{allworth2021transfer} Observations from the initial 20 minutes were rescaled to a uniform grid consisting of 1,200 points.
    \item \textbf{FURFARO}~\cite{furfaro2018space} - The first 500 points were utilized as input.
    \item \textbf{YAO}~\cite{yao2021basic} - The light curve was folded with the apparent rotation period to a grid with 200 points.
\end{itemize}

\begin{table*}[htbp]
    \centering
    \caption{Parameters used in the training process for each model.}
    \label{tab:params}
    \begin{tabular}{c|c|c|c|c}
    \toprule
         \multicolumn{1}{c}{} & \textbf{Input size} & \textbf{Channels} &  \textbf{Batch size} & \textbf{Scheduler} \\ \midrule
         \textbf{Default} & 10 000 & Mag &  32 &  x \\ 
         \textbf{ALLWORTH\cite{allworth2021transfer}} & 1 200 & Mag + Phase & 256 & x \\ 
         \textbf{ResNet\cite{he2016deep}} & 10 000 & Mag &  32 & x \\ 
         \textbf{FURFARO\cite{furfaro2018space} } & 500 & Mag &  32 & x \\ 
         \textbf{YAO\cite{yao2021basic}  } & 200 & Mag &  32 & x \\ 
         \textbf{Astroconf.\cite{pan2024astroconformer}} & 10 000 & Mag  & 32 & \checkmark \\ 
         \bottomrule

    \end{tabular}
\end{table*}

\begin{table*}[htbp]
    \centering
    \caption{Comparison of the Allworth method\cite{allworth2021transfer} on our dataset and the original paper.}
    \label{tab:allworth}
    \begin{tabular}{c|c|c|c|c|c|c}
\toprule
\multicolumn{1}{c|}{\textit{\textbf{Rocket Bodies}}} & \multicolumn{2}{|c|}{\textbf{F1 score}} & \multicolumn{2}{|c|}{\textbf{Precision}} & \multicolumn{2}{|c|}{\textbf{Recall}} \\
 \multicolumn{1}{c|}{} & \textbf{Our} & \textbf{Original}  & \textbf{Our} & \textbf{Original}  & \textbf{Our} & \textbf{Original} \\ 
\midrule
Atlas 5 Centaur & 0.39 & 0.74 & 0.53 & 0.71 & 0.33 & 0.78 \\
CZ-3B & 0.65 & 0.63 & 0.77 & 0.61 & 0.57 &  0.65 \\
Delta 4 & 0.11 & 0.81 & 0.19 & 0.78 & 0.09 & 0.85  \\
Falcon 9 & 0.49 & 0.80 & 0.48 & 0.81 & 0.49 & 0.79 \\
H-2A & 0.56 & 0.79 & 0.46 & 0.82 & 0.72 & 0.77 \\
\bottomrule
    \end{tabular}
    
\end{table*}

Only one paper offers a basis for comparison. We trained the Allworth method~\cite{allworth2021transfer} on our dataset and compared the results with those reported in the original publication. However, the comparison is not entirely fair, as the training scenarios differ. Specifically, the authors in~\cite{allworth2021transfer} used a balanced dataset with only 500 samples per class. Table \ref{tab:allworth} shows the result for the overlapping classes between the two datasets. The source code for the experiments is publicly available at \url{https://github.com/kyselica12/RoBo6_Model_Comparison}.




\section{USE CASE II: SURFACE CHARACTERIZATION OF ATLAS 2AS ROCKET BODY}\label{sec:sur_char} 

The constantly increasing population of artificial space objects, including satellites and space debris, leads to unwanted negative effects on the surrounding environment. The affected areas include the space near Earth \cite{PARDINI2011557}, the environment around the Moon \cite{Campbell2023}, Earth's atmosphere \cite{bartkova2022artificial}, and the ground \cite{Kocifaj2021}. Characterizing artificial space objects is a crucial methodology for accurately assessing the negative impact of specific objects or groups of objects, such as satellites in large mega-constellations or clusters of spared upper stages.
The analysis of photometric data published in photometric catalogues, such as MMT, can be used for phase function reconstruction \cite{silha2023ioc}. This process allows for a better understanding of the photometric properties of objects, aiding in brightness prediction as well as the physical characterization of an object. The phase function represents the dependence of an object's apparent brightness on the phase angle at which the data was acquired. This function contains physical information about the object's surface properties, such as porosity, reflectivity, and other characteristics.

As an initial analysis, especially for objects of unknown origin and shape, it is usually assumed that the object has a spherical shape with both diffuse and specular components present on its surface, both of which contribute to the overall apparent brightness detected by the observer. In their work \cite{MCCUE1971851}, the authors introduced the mixing parameter $\beta$, which allows the phase function to be modelled as a mixture of diffuse and specular reflections for a sphere. When $\beta$ = 1.0, the object's surface exhibits purely diffuse reflection, whereas, for $\beta$ = 0.0, the surface exhibits fully specular reflection.
For objects in the near-Earth environment, the following equation is often used to calculate the apparent brightness $m_V$, reduced to a range of 1000 km and a given phase angle $\phi$:


\begin{equation}
\label{eq_phase_function_diff_spec_1au}
    \begin{array}{r c l}
       m_V(1000km,\phi)=-26.74+5\text{log}(R_{km}) \\
       -2.5\text{log}(A\rho[\beta F_1(\phi)+(1-\beta)F_2(\phi)]),
    \end{array}
\end{equation}
where $R_{km}$ is the range between object and the observer [$km$] and functions $F_1(\phi)$ and $F_2(\phi)$ are diffuse and specular sphere phase functions respectively (see \cite{MCCUE1971851}). Coefficients $A$~[$m^2$] and $\rho$~[-] are mean cross-section and geometric albedo of the object respectively, parameters describing the physical characteristic of the object. 

For this analysis we used the LCDC tool to extract light curves belonging to the spared upper stage ATLAS 2AS CENTAUR R/B with NORAD ID 26858 from \texttt{MMT\_snapshot} dataset with the following preprocessing methods used in order: \texttt{FilterByPeriodicity} as we are only interested in \textit{"periodic"} samples, \texttt{SplitByRotationalPeriod} to divide samples into individual rotational periods, \texttt{FilterMinLength} with $\text{length}=100$ following by \texttt{FilterFolded} with $k=100$ and $\text{threshold}=0.9$ to filter out low-quality samples. Moreover, a couple of statistics were computed namely \texttt{MediumTime}, \texttt{MediumPhase} and \texttt{FourierSeries} of the $6^{th}$ order.

\begin{figure}[h!]
    \centering
    \includegraphics[width=7cm]{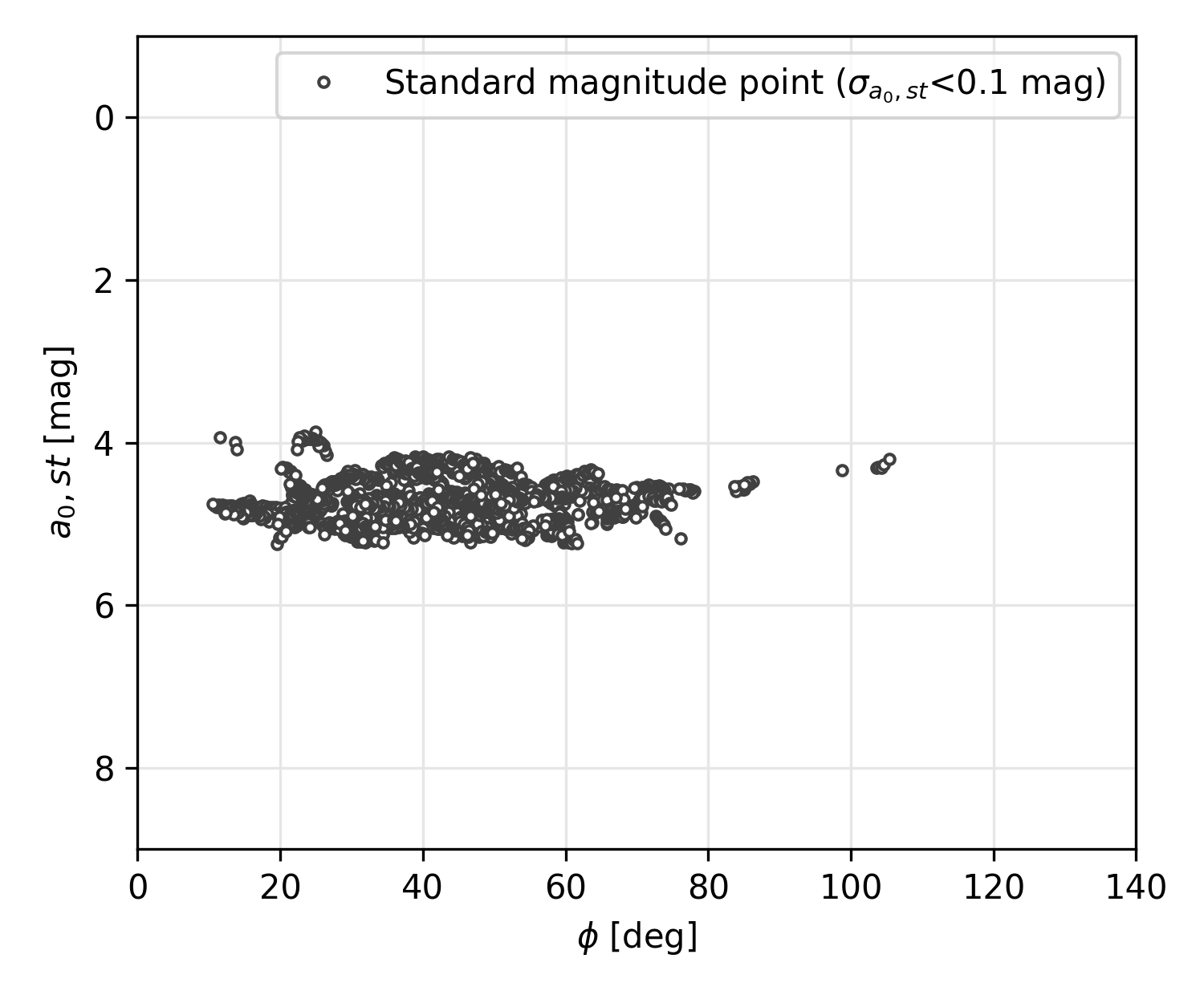}
    \caption{Coefficient $a_{0,st}$ calculated from standard magnitudes of object ATLAS 2AS CENTAUR R/B as a function of phase angle.}
    \label{fig:stMagnA0_Atlas2}
\end{figure}

\begin{figure}[h!]
    \centering
    \includegraphics[width=7cm]{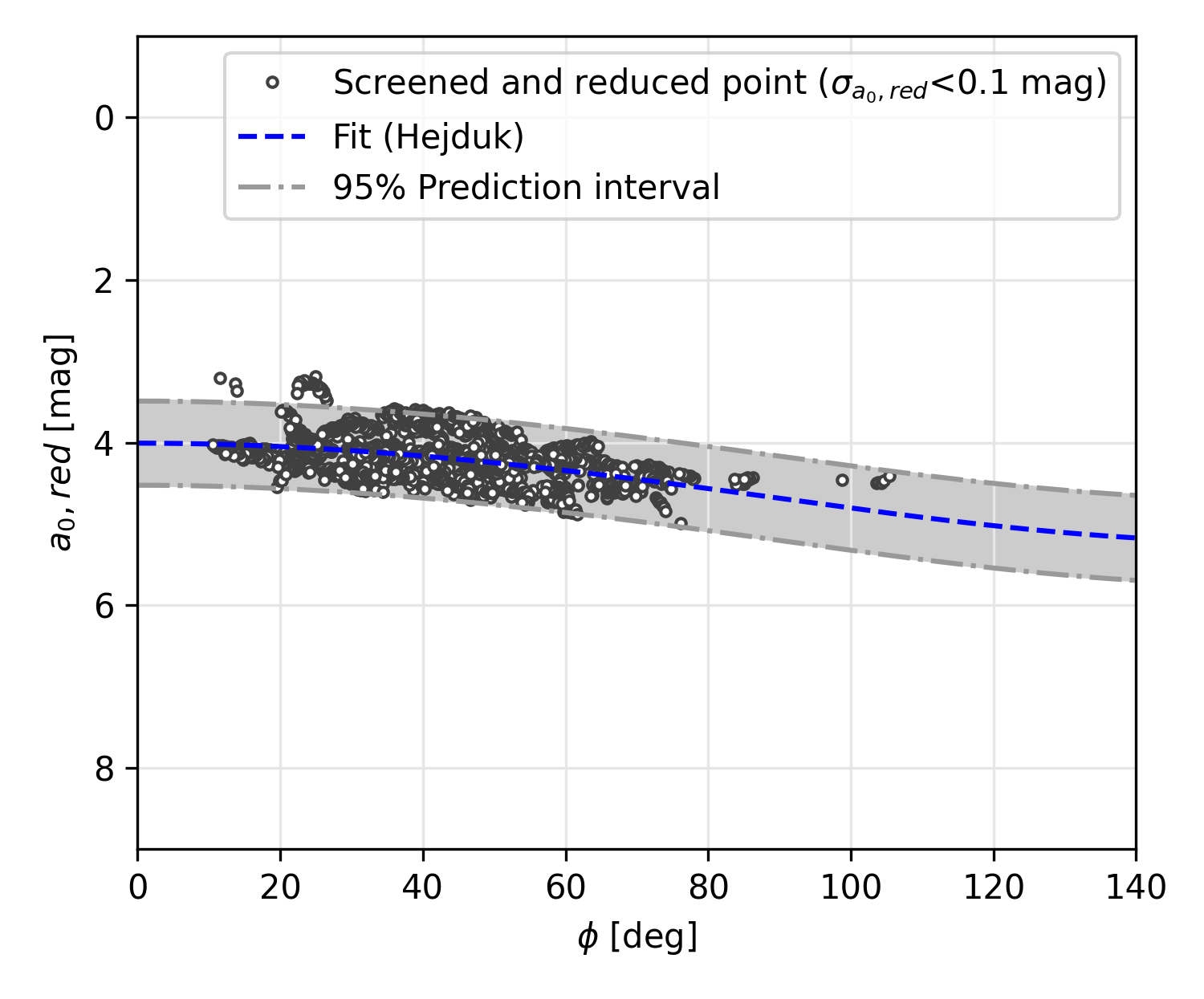}
    \caption{Coefficient $a_{0,red}$ calculated from $a_{0,st}$ coefficients of object ATLAS 2AS CENTAUR R/B as a function of phase angle (grey points). Plotted is the fit assuming a diffuse/specular sphere (blue dashed line) and 95$\%$ prediction interval (grey area).}
    \label{fig:fitPhaseFunAtlas2}
\end{figure}

In Figure~\ref{fig:stMagnA0_Atlas2} are plotted Fourier coefficients $a_{0,st}$ as a function of phase angle. 
Standard magnitude is defined according to \cite{karpov2016massive}, as apparent magnitude reduced to a range of 1000~$km$ and $\phi$ = 90° degrees. To extract the relevant physical parameters of the object, we removed the reduction to $\phi$ = 90° and we obtained the apparent magnitude values $a_{0,red}$ reduced only to a range of 1000~$km$ (Figure~\ref{fig:fitPhaseFunAtlas2}).

Once reduced, the data were fitted using equation for diffuse/specular sphere (see Equation~\ref{eq_phase_function_diff_spec_1au}) and extracted were parameters $A\rho$ = 3.63$\pm$0.09~$m^2$, $\beta$=0.44$\pm$0.03 and absolute magnitude $H_{1000km}$=4.0$\pm1.5$~$mag$. The so-called prediction interval was also calculated, defining the area with a 95$\%$ probability of containing the given brightness. Parameter $\beta$ indicates the presence of highly reflective features on the object's surface, which is a strong indication of artificial materials. The coefficient $A\rho$ indicates larger object, with mean cross section of 20.7~$m^2$ and resulting diameter of $d$ = 5.1~$m$ assuming $\rho$ = 0.175 or 17.5~$\%$ \cite{Mulrooney2008}.
Estimated values together with Equation.~\ref{eq_phase_function_diff_spec_1au} can be further used for object brightness prediction to support its detectability and cataloguing with optical telescopes. 

\section{USE CASE III: EVOLUTION OF ROTATIONAL DYNAMICS OF DELTA 4 ROCKET BODY} \label{sec:rot_prop}

To stabilize the population of space debris, it will be necessary to either remove specific objects or extend their operational lifespan. Both of these processes require knowledge of the rotational properties of the target objects and the ability to predict future values of these parameters. This can be achieved by analysing the past evolution of rotational properties. The tool LCDC enables the determination of rotational period values for specific objects or a particular population. From these values, it is further possible to calculate angular velocity and the rate of change in the rotational period.

\begin{figure}[htbp]
    \centering
    \includegraphics[width=5cm]{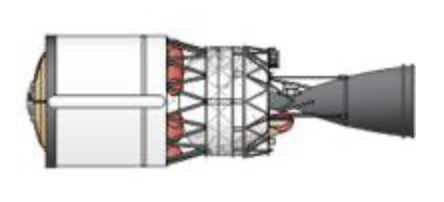}
    \caption{DELTA 4 rocket body \cite{historicspacecraft_upper_stages}.}
    \label{fig:delta}
\end{figure}

We used the LCDC to identify curves with the estimated period of the object DELTA 4, 40535, 2015-013B. For preprocessing we used \texttt{FilterByPeriodicity} to filter out non-variable and aperiodic samples, \texttt{FilterFolded} with parameters $\texttt{k}=100$ and $\texttt{threshold}=0.95$ following \texttt{FilterMinLength} with $\texttt{length}=100$ to filter out low-quality samples. The object is shown in Figure~\ref{fig:delta}. It is a cylindrical rocket body orbiting the Earth on a nearly circular trajectory at an altitude of approximately 20,900 km \cite{n2yo_40535}. Using the tool, we were able to retrieve all observation dates and the corresponding rotational periods of this object. From these values, we determined the rotation period change $\alpha$ using the Equation~\ref{eq:rotation_change},

\begin{equation}
\alpha=\frac{P_2-P_1}{T_2-T_1},
\label{eq:rotation_change}
\end{equation}
where \( T_2 \) is the time of the later observation, \( P_2 \) is the rotational period at that time, \( T_1 \) is the time of the earlier observation, and \( P_1 \) is the rotational period at that time.

Angular velocity is calculated by the Equation~\ref{eq:angular},

\begin{equation}
\omega=\frac{360^\circ}{P},
\label{eq:angular}
\end{equation}
where $P$ is the rotation period of the object.

\begin{figure}[htbp]
    \centering
    \includegraphics[width=\linewidth]{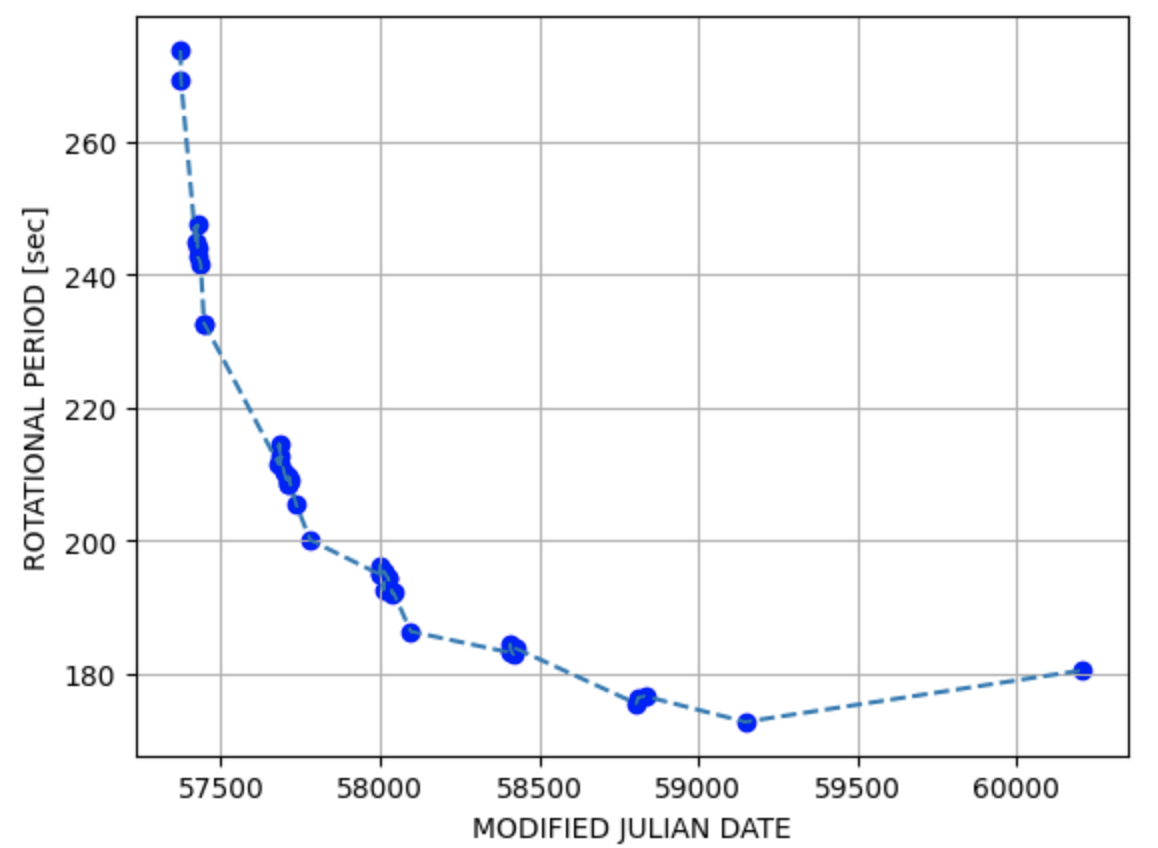}
    \caption{Evolution of rotation period of DELTA 4, 40535, 2015-013B.}
    \label{fig:rotation_period}
\end{figure}

\begin{figure}[htbp]
    \centering
    \includegraphics[width=\linewidth]{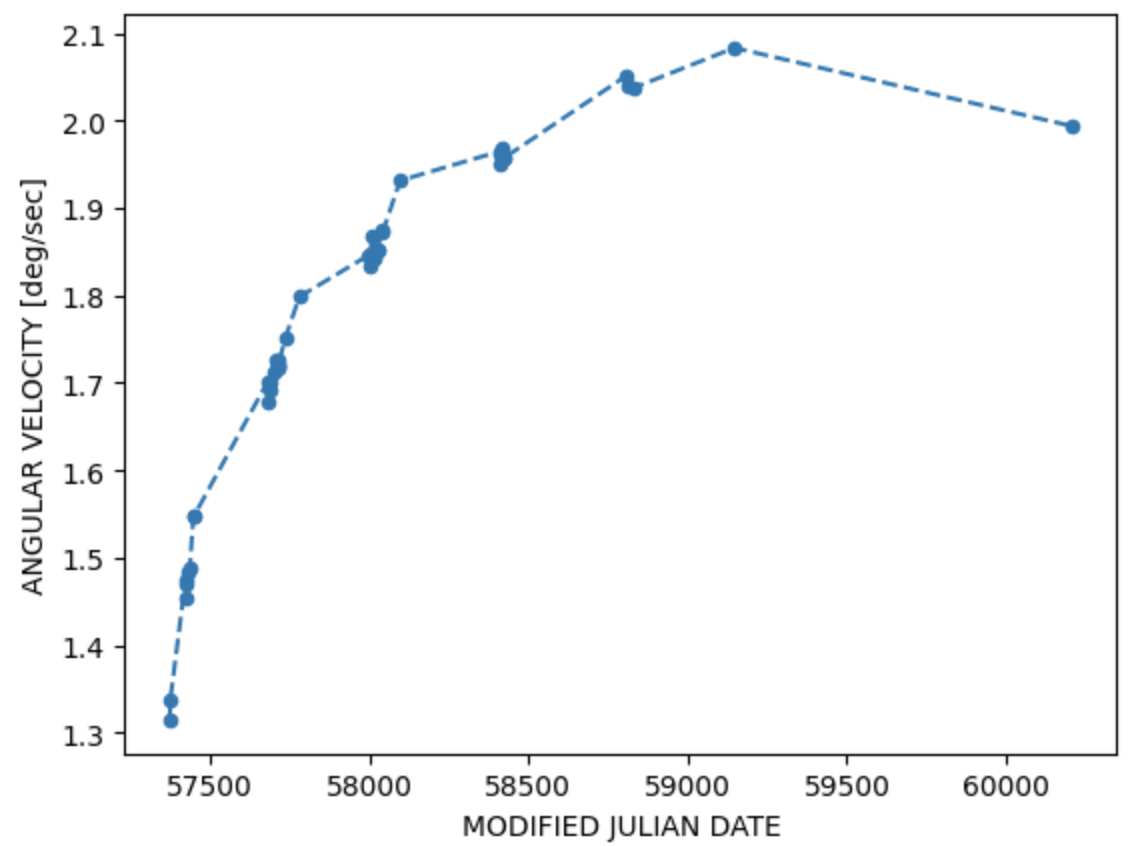}
    \caption{Evolution of angular velocity of DELTA 4, 40535, 2015-013B.}
    \label{fig:angular_velocity}
\end{figure}

\begin{figure}[h!]
    \centering
    \includegraphics[width=\linewidth]{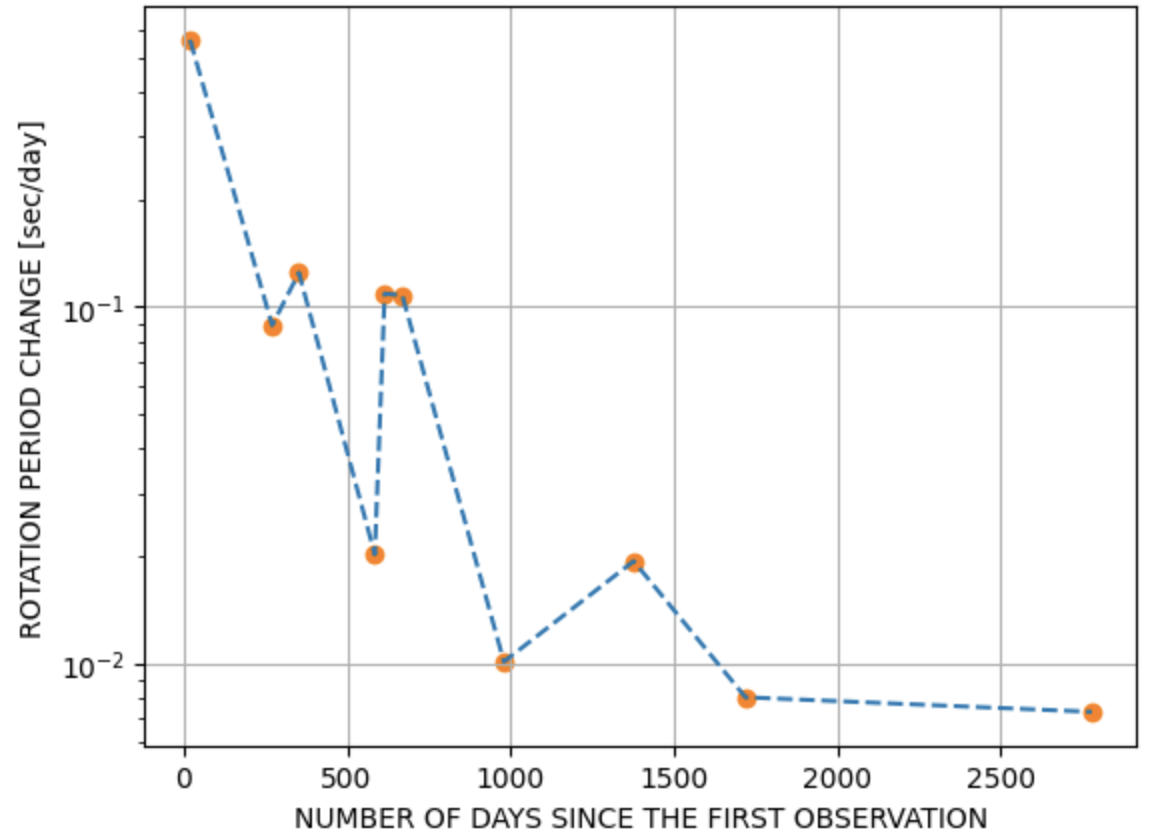}
    \caption{Evolution of rotation period change of DELTA 4, 40535, 2015-013B.}
    \label{fig:rotation_period_change}
\end{figure}

As seen in Figure~\ref{fig:rotation_period}, the initial rotational period is 273 seconds. This value is from December 2015. Subsequently, the rotational period changes more rapidly for about 2 months and more slowly for approximately 8 months. Such cycles repeat a total of three times. The rotational period then decreases by 10 seconds over approximately 700 days. The most recent data is from September 2023, when the rotation period was about 180 seconds. Thus, the rotation period has increased.

In Figure~\ref{fig:rotation_period_change}, the rate of change of the rotation period in seconds per day is displayed. Here, it is evident that this rate varies during individual cycles but has overall decreased by two orders of magnitude. Towards the end, the rotational period changes at a rate of hundredths of a second per day. The angular velocity, shown in Figure~\ref{fig:angular_velocity} changed over the entire period from 1.3 $^\circ$/sec to 2.1 $^\circ$/sec. 

All the described changes suggest that the object was influenced by solar radiation pressure, which can cause quasi-annual variations in the rotation period. This influence might be dominant, as the object is in an orbit where the atmospheric drag and magnetic field should no longer have a significant effect \cite{earl2015observations}. These data can be further used as input for simulating rotational properties.

\section{CONCLUSIONS}

This paper introduces the Light Curve Dataset Creator (LCDC), a versatile Python-based toolkit for preprocessing, analyzing, and generating datasets from space debris light curve data, specifically tailored for applications in machine learning and astronomy. By providing tools to handle large datasets, including the newly introduced \texttt{MMT\_snapshot} dataset, LCDC addresses significant challenges in accessibility and reproducibility that have hindered prior research.

To demonstrate the capabilities of the toolkit, we propose RoBo6, the first standardized dataset for rocket body classification, enabling consistent evaluation and comparison of machine learning models. Benchmarks on RoBo6 confirmed its suitability for training state-of-the-art models, including transformer-based architectures, and highlighted the advantages of a standardized approach to dataset preparation and evaluation.

Furthermore, we showcased the utility of LCDC in two scientific use cases: determining surface properties of the Atlas 2AS Centaur upper stage and analyzing the rotational dynamics evolution of the Delta 4 rocket body. These examples emphasize how LCDC simplifies the preprocessing and analysis of light curve data, enabling deeper insights into the physical and rotational characteristics of space objects.

By addressing the limitations of existing tools and datasets, LCDC represents a significant step forward in advancing research in space debris characterization and sustainable space exploration. Its open-source nature ensures that the scientific community can benefit from its capabilities and build upon this foundation to further innovate in the field.

\section*{ACKNOWLEDGEMENTS}
Mini-MegaTORTORA belongs to Kazan Federal University.

\bibliographystyle{ieeetr}
\bibliography{references}

\end{document}